\documentclass[utf8]{FrontiersinHarvard}

\usepackage{url,hyperref,lineno,microtype,subcaption} 
\usepackage[onehalfspacing]{setspace}
\usepackage{soul}
\usepackage{booktabs}
\usepackage{siunitx} 
\usepackage{amsmath}
\usepackage{chngcntr}   

\def\keyFont{\fontsize{8}{11}\helveticabold }
\def\firstAuthorLast{Jafarzadeh {et~al.}}
\def\Authors{Shahin Jafarzadeh\,$^{1,*}$, David B. Jess\,$^{1,2}$, Marco Stangalini\,$^{3}$, Peter H. Keys\,$^{1}$, Glen Chambers\,$^{1}$, Samuel D. T. Grant\,$^{1}$, Michele Berretti\,$^{4,5}$, and Timothy~J.~Duckenfield\,$^{1}$}

\def\Address{
$^{1}$Astrophysics Research Centre, School of Mathematics and Physics, Queen’s University Belfast, Belfast, BT7 1NN, Northern Ireland, UK \\
$^{2}$Department of Physics and Astronomy, California State University Northridge, Northridge, CA 91330, USA \\ 
$^{3}$ASI Italian Space Agency, Via del Politecnico snc, 00133 Rome, Italy \\
$^{4}$University of Trento, Via Calepina 14, 38122 Trento, Italy \\
$^{5}$University of Rome Tor Vergata, Department of Physics, Via della Ricerca Scientifica 3, 00133 Rome, Italy
}

\def\corrAuthor{Shahin Jafarzadeh}

\def\corrEmail{shahin.jafarzadeh@qub.ac.uk}

\begin{document}
\onecolumn
\firstpage{1}

\title[Adaptive multi-line fitting]{Adaptive multi-line fitting for stable line-core intensity and Doppler velocity} 
\author[\firstAuthorLast ]{\Authors} 
\address{} 
\correspondance{} 
\extraAuth{}

\maketitle

\begin{abstract}
Next-generation solar spectrographs increasingly record dense wavelength windows in which tens to hundreds of spectral lines are sampled at each spatial location and time step. This expands the scope for multi-line, multi-height diagnostics of magnetohydrodynamic (MHD) motions, but also raises a practical challenge: deriving stable line-core intensity and line-of-sight (LOS) velocity time series when profiles evolve rapidly, become asymmetric, blend, or develop multi-lobed cores. Common fast estimators (e.g. parabolic minima, centre-of-gravity measures, Fourier-phase proxies, and fixed-window symmetric fits) can perform well for simple, isolated absorption lines, yet can intermittently misidentify the core in crowded or morphologically complex cases. Even infrequent mis-tracking can leave step-like artefacts that redistribute power and bias spectral, phase, and coherence measures used in wave and dynamics analyses.
We introduce \texttt{LineFit}, a fully reproducible adaptive multi-line fitting approach tailored to dense-window spectroscopy from facilities such as DKIST, {\sc Sunrise iii}, and integral-field instruments. \texttt{LineFit} models each line locally with bounded non-linear least-squares fits to a Voigt-family profile, including an asymmetric-Voigt option to accommodate unequal wing broadening, and incorporates close-pair ownership control together with conservative, per-line window adaptation and split-core-aware handling. Using a synthetic near-UV time series with unambiguous ground truth, we benchmark \texttt{LineFit} against four widely used fast baselines and assess both instantaneous centre errors and downstream time-series diagnostics. Several fast methods remain competitive for many lines, whereas \texttt{LineFit} is most robust in key stress cases involving intermittently split-core profiles and correspondingly yields power spectra that agree most closely with the truth. We also demonstrate a proof-of-principle that benchmarks hybrid acceleration of the \texttt{LineFit} software via supervised emulation, offering at least three orders-of-magnitude improvement in processing time, with 87\% of validation samples agreeing with \texttt{LineFit} to within 0.1\,pm in recovered line-centre position. All code and notebooks required to reproduce the testbed, figures, and results are publicly available, providing a portable benchmark and a practical basis for stable multi-line time-series extraction, with a clear route towards hybrid acceleration via supervised emulation.

\tiny
\keyFont{ \section{Keywords:} The Sun, MHD, Waves, Oscillations, Spectroscopy, Methods} 
\end{abstract}

\section{Introduction}

New flagship facilities and modern spectroscopic designs are shifting the practical limiting factor in wave and dynamics studies of the solar atmosphere \citep{2023LRSP...20....1J}. Instruments now routinely provide high spatial resolution, high cadence, and dense spectral sampling, increasingly in configurations that deliver many spectra per exposure \citep[e.g. integral-field units (IFUs) and multiplexed/slit-based spectrographs;][]{2019OptEn..58h2417I, 2020SPIE11447E..0YK, 2021SoPh..296...70R, 2025SoPh..300...65F, solankietal2026}. In this regime, dense spectral windows containing tens to hundreds of lines can be recorded for every spatial sample and time step, enabling multi-line constraints on magnetohydrodynamic (MHD) motions across atmospheric layers. At the same time, they can create an underappreciated bottleneck: converting raw spectra into science-grade, stable line-core intensity and line-of-sight (LOS) velocity time series at scale. This bottleneck is particularly relevant for next-generation datasets in which the information content is high but profile morphology is not static in the highly dynamic solar atmosphere, and where accurate time-series extraction must remain reliable both for well-behaved profiles and during episodes of strong profile distortion.

The scientific motivation for reliable line-core time series is broad. Doppler shifts and line-core intensity variations are key observables in studies of MHD waves and oscillations, shocks, flows, and small-scale transients, and they underpin frequency-phase-coherence analyses, wave-mode identification, and multi-height coupling inferences \citep[e.g.][]{2007A&A...463.1153T, 2012RSPTA.370.3193D, 2015SSRv..190..103J, 2021A&A...649A.169S, 2022NatCo..13..479S, 2022ApJ...938..143G, 2023A&A...674A.109C, 2024ApJ...970...66B, 2025A&A...697A.156B, 2025ApJ...982..202D}. In dense-window spectroscopy, their value can increase because multiple diagnostics can be extracted simultaneously from different lines sampling different formation regimes. However, the same dense windows may also intensify algorithmic challenges, particularly if blends become common, line widths vary strongly across the field of view and in time, asymmetries appear and disappear, or some lines develop multi-lobed cores or self-reversals. In such conditions, even occasional misidentification of the line core can be scientifically consequential because it can introduce discontinuities that inject broadband artefacts, bias power spectra, and degrade phase and coherence measurements.

A wide range of techniques are used in practice to infer LOS velocities and related atmospheric parameters. At one end, fast proxies estimate a representative line position from the observed intensity profile within a fixed or semi-fixed window: e.g. parabolic interpolation around the minimum, centre-of-gravity (COG) estimators of the absorption depression, bisectors, Fourier/phase-based centroids (Fourier-tachometer style proxies), polynomial core fits (including weighted variants), or symmetric parametric fits (Gaussian/Lorentzian) in a prescribed window \citep[e.g.][]{2003ApJ...592.1225U, 2012SoPh..278..217C, 2013ASPC..478...93B, 2021RSPTA.37900172G}. These approaches can be accurate and efficient for isolated, near-symmetric profiles with stable continua, and they often remain the only feasible option when computational cost limits what can be processed. Their limitations are also well known: if neighbouring features or blends fall within the chosen wavelength window, they can bias methods that compute the centre from the full windowed profile (e.g. COG or Fourier-phase centroids); asymmetric or evolving wings can systematically shift COG-like measures; and multi-lobed cores can cause the inferred ``centre'' to hop between competing minima or lobes, producing discontinuities in time series that are difficult to identify and correct in subsequent analysis \citep{2001A&A...374..733B, 2013A&A...557A..93F, 2018ASSP...49..181F}.

At the other end, spectropolarimetric inversions infer atmospheric parameters by fitting Stokes profiles under radiative-transfer constraints \citep[][]{1992ApJ...398..375R}. Inversions (and related forward-modelling frameworks) provide a physically interpretable route to LOS velocities (and magnetic/thermodynamic structure) and can be essential when magnetic diagnostics are measured. They also come with substantial computational cost, model assumptions, and practical sensitivity to initialisation, parameterisation, and data quality. Inversion families span Milne–Eddington \citep[e.g.][]{2004A&A...414.1109L, 2007MmSAI..78..148L, 2011SoPh..273..267B} and depth-dependent approaches \citep[e.g.][]{2000A&A...358.1109F, 2015A&A...577A...7S, 2018A&A...617A..24M}, LTE\footnote{Local Thermodynamic Equilibrium} and non-LTE formalisms, and increasingly multi-line strategies that exploit complementary constraints from multiple transitions \citep[e.g.][]{2019A&A...623A..74D, 2019A&A...622A..36R, 2022A&A...660A..37R, 2026A&A...705A.220H}. In practice, the choice of inversion method and configuration depends on the scientific objective, the line(s) observed, and the cadence/volume constraints, and it is often not feasible to run full inversions for every pixel and time step in very large field-of-view and high-cadence datasets. As a result, many workflows still rely on line-core time-series extraction as an intermediate, scalable product even when inversions are applied to subsets or representative regions.

Between these extremes, machine-learning (ML) approaches and emulators are increasingly used to accelerate profile interpretation and/or produce robust proxies. Examples include supervised tools designed to fit or classify chromospheric line profiles in an automated manner (e.g. the MCALF framework; \citealt{2021RSPTA.37900171M}), as well as broader efforts to emulate expensive inversions or to enable robust, scalable classification of spectropolarimetric profile morphologies for large surveys and statistical comparisons \citep{2019A&A...626A.102A, 2025ApJ...985L...7J, 2025ApJ...988....9C}. While ML strategies can deliver dramatic speed-ups, their reliability depends on training coverage, careful validation, and explicit handling of out-of-distribution profiles -- challenges that become important precisely in the highly dynamic regimes, such as those for wave and transient studies. Irrespective of whether future workflows rely on parametric fitting, ML, or hybrids, well-characterised reference extractions provide a necessary anchor for validation, uncertainty assessment, and quality control in these dynamic regimes.

In this paper we focus on a complementary but widely needed product: robust, automated extraction of line-core intensity and LOS velocity time series across many lines in dense windows, suitable for subsequent wave and dynamics analysis. Our goal is not to replace inversions, but to provide an adaptive, reproducible approach that reduces failure modes of common fast estimators in crowded and evolving spectra, and that scales to multi-line use cases. As such, this method-focused contribution is intended to be immediately usable by observers working with, e.g., Daniel K. Inouye Solar Telescope (DKIST; \citealt{2016AN....337.1064T}) and {\sc Sunrise iii} balloon-borne solar observatory \citep{2025SoPh..300...75K} spectroscopy, and Fibre Resolved OpticAl and Near-Ultraviolet Czerny-Turner Imaging Spectropolarimeter (FRANCIS; \citealt{2023SoPh..298..146J}), and it is readily transferable to other slit and IFU spectrographs that deliver (or will deliver) time-dependent line profiles \citep[e.g., the Interface Region Imaging Spectrograph (IRIS), the Multi-slit Solar Explorer (MUSE), and Solar-C;][]{2014SoPh..289.2733D, 2020ApJ...888....3D, 2021SPIE11444E..0NS}, while also remaining applicable to single-line time-series studies. We present \texttt{LineFit}, an adaptive multi-line fitting methodology built around four design principles: (i) line-guided seeding near expected centres (with explicit handling of close blended pairs), (ii) conservative per-line adaptive windowing with hard safety bounds, (iii) iterative centre refinement for stability under profile evolution, and (iv) optional wing-weighted or wing-masked fitting to suppress lobe-hopping when the core becomes complex. We benchmark the method using a realistic synthetic near-UV testbed with unambiguous ground truth, compare against a number of widely used fast baselines, and quantify not only per-time-step centre errors but also the impact on standard time-series wave products. Finally, we briefly note how such a robust fitting baseline could support scalable processing strategies, including hybrid acceleration via supervised emulation.

\section{Materials and Methods}
\label{sec:data_methods}

This section describes the validation dataset and the comparison baselines used throughout the paper. We adopt a synthetic, dense-window spectral time series because it provides unambiguous ground truth for the line-centre evolution and enables controlled stress tests (blends, asymmetries, close pairs, and split/reversal morphologies) that are difficult to isolate and quantify in real observations. The goal is not to replace real-data demonstrations, but to establish a reproducible performance benchmark where centre errors and time-series artefacts can be measured directly. All routines used to generate the synthetic dataset, run the estimators, and reproduce all figures presented in this work are publicly available in a GitHub repository described in Section~\ref{subsec:code_availability}.

\subsection{Synthetic multi-line testbed}
\label{subsec:testbed}

We validate and benchmark the proposed adaptive multi-line fitting methodology using a realistic synthetic near-UV spectral time series designed to mimic modern dense-window spectroscopy, where tens of absorption lines coexist within a narrow wavelength interval, and line profiles evolve significantly in time. The testbed consists of $N_{\mathrm{lines}}=10$ diagnostic lines sampled on a common wavelength grid $\lambda\in[\lambda_{\min},\lambda_{\max}]$, spanning a total window width $\Delta\lambda\equiv\lambda_{\max}-\lambda_{\min}\approx 0.36$\,nm around a central wavelength of approximately 299.90\,nm, with $N_\lambda = 2400$ spectral samples (uniform spacing $d\lambda = (\lambda_{\max}-\lambda_{\min})/(N_\lambda-1)\approx 1.5\times10^{-4}$\,nm).
The cadence is $\Delta t = 10$\,s and the duration is $N_t = 300$ time steps ($T=3000$\,s). Each time step contains a full spectrum $I(\lambda, t)$ in which (i) line depths and widths vary smoothly in time, (ii) asymmetric profiles exhibit time-dependent core skewness, (iii) a close blended pair has controlled separation, and (iv) one line intermittently develops a split (two absorption lobes) with a central emission-reversal component. In addition, weak blending contaminants are injected near selected lines to mimic crowding in dense windows.

All line profiles are synthesised from Voigt-based primitives. For a nominal rest wavelength $\lambda_{0,i}$, the time-dependent Doppler-shifted line-centre wavelength is,
\begin{equation}
\lambda_i(t) = \lambda_{0,i}\left(1+\frac{v_i(t)}{c}\right) \ ,
\end{equation}
where $\lambda_i(t)$ is the ground-truth line-centre wavelength of line $i$ at time $t$, $v_i(t)$ is the imposed LOS velocity driver and $c$ is the speed of light. The absorption component of each line is represented by a normalised Voigt shape with time-dependent depth and widths. Asymmetric profiles are produced by a controlled, time-dependent core skewness (i.e. the deepest point is displaced relative to the nominal Doppler-shifted centre), so that local-minimum based estimators face a realistic ambiguity in ``what constitutes the core''. The split+reversal morphology is produced by transitioning between a single absorption core and a two-lobe split profile with time-dependent separation $\Delta\lambda_{\mathrm{split}}(t)$, and by adding a Gaussian emission component localised near the nominal line centre. Importantly, the central emission feature is not required to exceed the continuum; throughout this work, ``emission reversal'' refers to a local core brightening relative to the surrounding absorption depression, i.e. a split/multi-lobed structure in the core region.

\subsubsection{Time-dependent evolution and truth products}
\label{subsec:truth}

Temporal evolution is imposed through stochastic-but-smooth processes that approximate non-stationary solar dynamics. For each line, the LOS velocity $v_i(t)$ is generated as a multi-frequency signal composed of three to five sinusoidal components with random phases, time-varying amplitudes, small time-dependent frequency drift, low-frequency jitter, and intermittent smoothed impulses. Component frequencies are drawn within 2--9\,mHz. To avoid artificial discontinuities, no hard clipping is applied to $v_i(t)$; instead, each $v_i(t)$ is smoothly rescaled by a single multiplicative factor so that $\max |v_i(t)|$ does not exceed a prescribed per-line cap. In the configuration used here, caps span 1.2--3.5\,km\,s$^{-1}$ across the ten lines. Line depth and width parameters evolve via first-order autoregressive [AR(1)] drivers, and the split+reversal line includes additional stochastic modulation of the splitting strength and core emission amplitude.

The synthetic FITS product stores, in addition to the spectrum cube $I(\lambda,t)$, the rest wavelengths $\lambda_{0,i}$, the ground-truth time-dependent line-centre wavelengths $\lambda_i(t)$, and the truth velocities $v_i(t)$. These truth products enable direct, per-time-step centre residuals for any estimator, defined as
\begin{equation}
\Delta\lambda_i(t)=\hat{\lambda}_i(t)-\lambda_i(t) \ ,
\end{equation}
where $\hat{\lambda}_i(t)$ is the estimated line-centre wavelength returned by the method.

\subsubsection{Noise and instrumental effects}
\label{subsec:noise}

To approximate realistic measurement conditions, the synthetic spectra include (i) a weak continuum gradient across the window, implemented in continuum-normalised (dimensionless) intensity units as
\begin{equation}
I_{\rm cont}(\lambda)=1+s\,\frac{\lambda-\lambda_{\rm mid}}{\Delta\lambda} \ ,
\end{equation}
where $I_{\rm cont}(\lambda)$ is the continuum level as a function of wavelength, $s$ is the dimensionless gradient amplitude, $\lambda_{\rm mid}\equiv(\lambda_{\min}+\lambda_{\max})/2$ is the midpoint of the simulated wavelength window, and $\Delta\lambda\equiv\lambda_{\max}-\lambda_{\min}$ is the full window width. In the benchmark presented here we use $s=0.02$, (ii) a low-amplitude quasi-periodic ripple to mimic residual fringing/throughput structure (amplitude $1.5\times10^{-3}$ and wavelength period of 0.035\,nm), and (iii) additive Gaussian noise whose amplitude increases in deep absorption cores (``core-boosted'' noise). In the configuration used here, the continuum noise standard deviation is $\sigma_{\rm cont}=2.1\times10^{-3}$ (relative intensity units) and the noise is increased in deep cores by a scaling boost factor of 2.0.

Optionally, a spectral point-spread function (PSF) convolution can be applied along the wavelength axis; however, we do not include PSF convolution in the benchmark presented here in order to isolate algorithmic centre-tracking behaviour from instrument-dependent spectral broadening and to keep the ground-truth morphology and blend structure as directly comparable as possible across methods (the PSF option remains available for sensitivity tests and instrument-specific tuning).

Other observational effects, such as scattered light, residual flat-field errors, spatially correlated noise, or imperfect wavelength calibration, are not explicitly included in the baseline benchmark presented here. These could modify the apparent core depth, local contrast, or wing shape, and therefore affect the absolute performance of any centre-tracking method. In general, methods that use constrained local fitting and conservative window selection are expected to be less sensitive to such complications than unconstrained fast estimators, although quantifying this in detail is beyond the scope of the present benchmark. In the case of \texttt{LineFit} (described in Section~\ref{subsec:walsalinefit}), the bounded fitting, conservative window adaptation, and morphology-aware handling are specifically intended to reduce sensitivity to such complications by preventing large excursions toward unrelated structures or unstable competing minima.

These perturbations are also visible in Figure~\ref{fig:fig1} as (i) a slight large-scale tilt of the local continuum across the full spectral window, produced by the weak continuum gradient, (ii) low-amplitude quasi-periodic waviness between and around lines, produced by the ripple term, and (iii) small local intensity irregularities that become more noticeable in deeper cores, produced by the core-boosted Gaussian noise.

\subsection{Adaptive multi-line profile fitting: \texttt{LineFit}}
\label{subsec:walsalinefit}

The proposed method (\texttt{LineFit}) is designed for dense spectral windows with many lines and evolving morphologies, where a single fixed fitting configuration is insufficient. In the implementation used here, each spectrum is first sanitised for non-finite samples (interpolated along wavelength index where needed), then robustly normalised to the local min--max range, lightly smoothed (optional), and intensity-inverted so that absorption cores become peaks
\footnote{Here ``inverted'' refers only to an intensity transformation (absorption minima mapped to peaks). It does not refer to Stokes inversions or atmospheric inversions.}. \texttt{LineFit} then enforces stable centre tracking via three coupled stages:

\begin{enumerate}
\renewcommand{\labelenumi}{(\roman{enumi})}
    \item \textit{Coarse seeding near expected centres (including close-pair ownership control).}
    Initial seeds are generated in the intensity-inverted spectrum by searching locally around each expected rest wavelength. Candidate peaks are identified via peak finding and scored using a (dominant) distance-to-prior term (preference for proximity to the expected centre) together with a (minor) peak-height term (a small preference for deeper, better-defined absorption cores, i.e. stronger peaks in the intensity-inverted spectrum). This means that shallow secondary valleys or unrelated nearby structures are not usually selected unless they remain sufficiently close to the expected centre, and for declared close pairs the explicit valley-based ownership split further prevents neighbouring lines from capturing each other's minima. Seeds are further guarded by a maximum allowed centre offset from the expected wavelength, preventing spurious matches to unrelated structures in crowded windows. For declared close pairs, ``ownership'' is enforced by splitting the shared search region at the inter-line valley, i.e. assigning each member of the pair its own side of the valley, so that the two members cannot swap under moderate profile evolution.

    \item \textit{Per-line adaptive windowing with hard safety bounds and morphology gating.}
    Each line $i$ starts from a per-line baseline half-window (or a global default), but the effective half-window is constrained by user-provided per-line minimum and maximum half-windows. Window adaptation is conservative: it expands only when justified by local measurements of profile width and complexity, and it can contract to protect the core when the measured depression width indicates that a smaller window is adequate (limited by the per-line minimum half-window). Two morphology-specific gates, i.e. decision rules that activate specialised handling only when a given profile morphology is detected, are applied in this paper: (a) for lines declared as potential self-reversal candidates, a split-core detector is evaluated time-step by time-step in the intensity-inverted profile; specifically, the detector searches for two sufficiently prominent local peaks separated by an inter-peak valley that drops below a prescribed fraction of the weaker peak, thereby distinguishing a genuine split-core structure from small fluctuations or noise. If such a morphology is detected, reversal handling is activated only for that time step, forcing the fit window to the permitted maximum so both lobes are contained. Operationally, this detector identifies a two-lobed core structure with a sufficiently deep inter-lobe valley in the intensity-inverted profile, so that the fit can be stabilised around the overall split-core morphology rather than jumping to one of the competing lobes; (b) for non-reversal lines, optional emission-bump and secondary-peak guards are used to avoid unsafe window growth in the presence of nearby competing structures (e.g. blends), thereby reducing the risk of window leakage. In practice, conservative thresholds are preferred: false positives mainly reduce computational efficiency by triggering safer handling more often than necessary, whereas false negatives are more consequential because they allow a genuinely complex profile to be treated too simply.

    \item \textit{Iterative refinement of centres with asymmetric handling.}
    Given the current window, a parameterised Voigt-family line model (Voigt or asymmetric Voigt) is fitted via bounded non-linear least squares optimisation in intensity-inverted space, with the fitted parameters defining the local core shape, width, amplitude, and centre. We adopt a Voigt-family model because it provides a flexible yet compact description that can accommodate Gaussian-like broadening (e.g. thermal/instrumental) together with Lorentzian-like wings (e.g. damping/pressure broadening), while remaining stable under bounded fitting. Purely Gaussian or purely Lorentzian fits can systematically misrepresent either the core or the wings for many photospheric/chromospheric lines. Our aim here is not to claim that the Voigt family provides a unique physical description of every observed profile, but rather to adopt a compact and sufficiently flexible parametric form that yields stable centre tracking across a wide range of morphologies. In this context, the centre position is the primary product of interest, and modest mismatch in the detailed wing shape is generally less important than avoiding unstable centre shifts between competing minima or lobes. The choice of a Voigt-family model should therefore be viewed as a pragmatic compromise between flexibility, stability, and computational cost, rather than as a full radiative-transfer description of the line formation process. The fitted centre is used to recentre the window and the fit is iterated for a fixed number of iterations or until convergence. Wing asymmetry is handled explicitly through an asymmetric Voigt parameterisation, allowing the fit to accommodate unequal wing broadening while maintaining bounded parameters. The parameter bounds are imposed as practical local constraints to keep the solution within a plausible regime for the current line and fitting window, rather than being derived from an external library of fixed Voigt parameters for each transition. In the present implementation, the choice between a Voigt and an asymmetric-Voigt profile is set by the fitting configuration, i.e. the algorithm does not automatically try both models and select between them at each time step. The refinement therefore contains two conceptually distinct levels: an outer loop in which the fitted centre is used to recentre the window and repeat the fit if needed, and an inner bounded non-linear least-squares optimisation that solves for the profile parameters within a given window. To further suppress discontinuities associated with lobe-hopping or unstable asymmetry solutions, the implementation includes (for asymmetric-Voigt fits) a guarded fall-back refit in which the asymmetry-parameter bounds are tightened when the initial solution lies close to parameter bounds or window edges. This fall-back is disabled for time steps where a split-core morphology has been detected, so that genuine multi-lobed behaviour is not artificially suppressed.
\end{enumerate}

\begin{figure*}[t!]
\centering
\includegraphics[width=\textwidth]{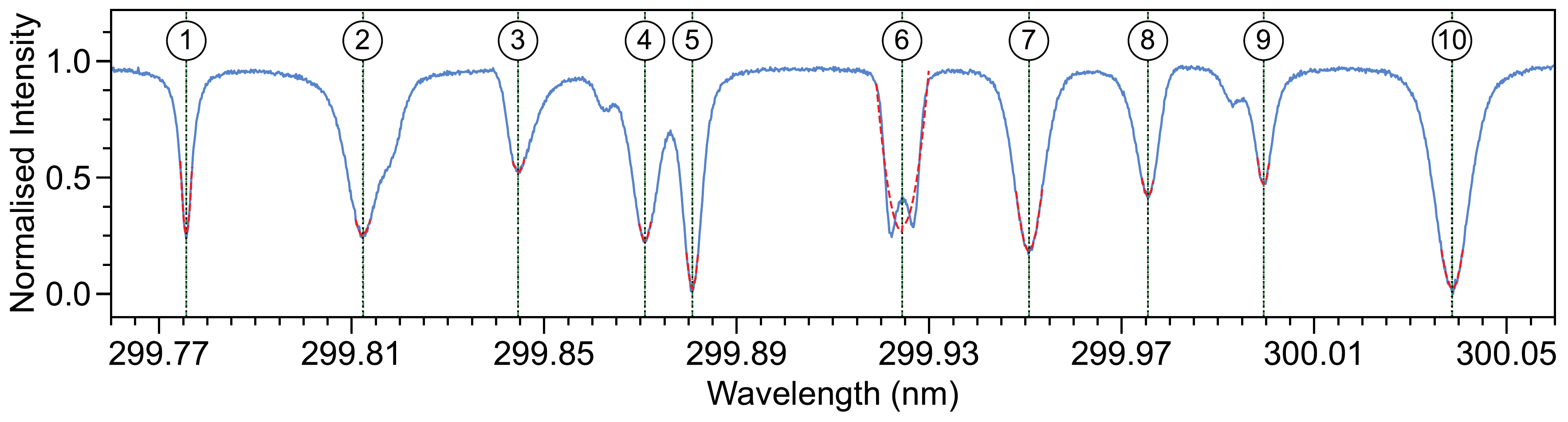}
\caption{Example synthetic spectrum and line-centre recovery. The blue curve shows the normalised spectrum; red dashed segments indicate the fitted model within each adaptive window. Vertical markers show the recovered centres (green dashed) and the truth centres (black dotted). Numbered circles label the ten diagnostic lines used throughout the testbed.}
\label{fig:fig1}
\end{figure*}

\begin{figure*}[h]
\centering
\includegraphics[width=\textwidth]{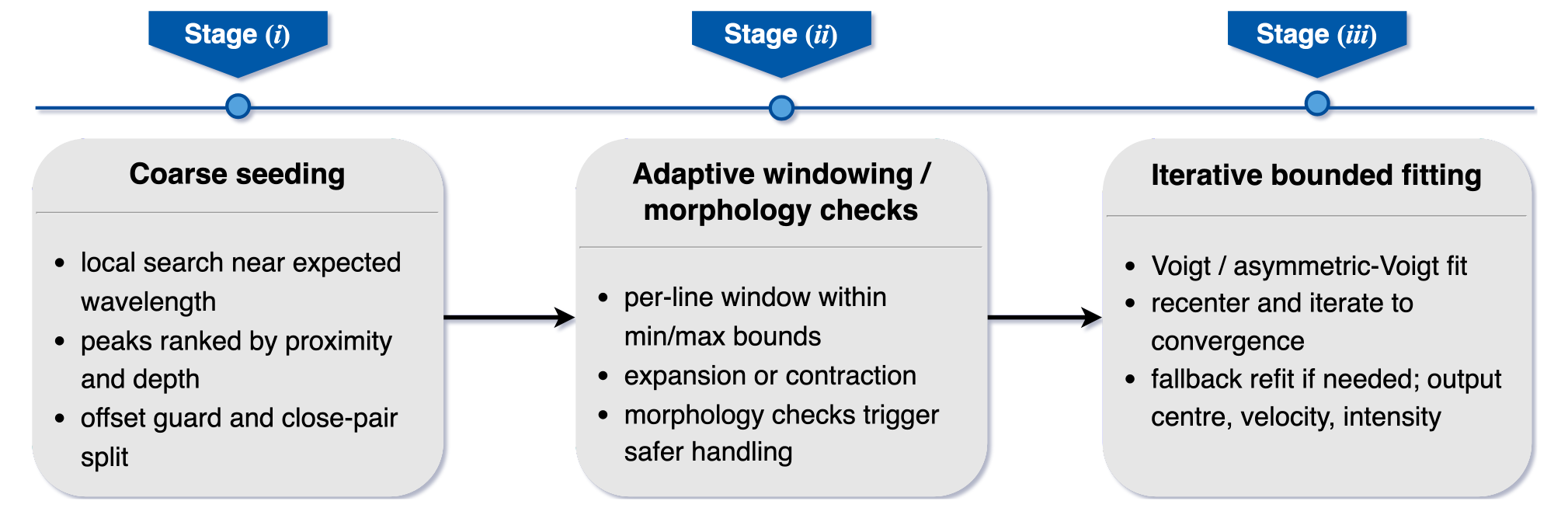}
\caption{Schematic overview of the \texttt{LineFit} workflow. The method proceeds through three coupled stages: (i) coarse seeding, including local search near the expected wavelength together with offset guarding and close-pair ownership control; (ii) adaptive windowing, in which per-line windows are conservatively adjusted within prescribed bounds and morphology checks activate safer handling when needed; and (iii) iterative bounded fitting, in which Voigt-family models are fitted, recentered, and iterated to convergence, yielding the recovered line centre, LOS velocity, and line-core intensity.}
\label{fig:linefit_workflow}
\end{figure*}

A schematic summary of this workflow is provided in Figure~\ref{fig:linefit_workflow}. The primary outputs are the fitted line-centre wavelengths $\hat{\lambda}_i(t)$, line-core intensity samples $I(\hat{\lambda}_i,t)$ evaluated at the fitted centre, and LOS velocities $\hat{v}_i(t)=c\,(\hat{\lambda}_i-\lambda_{0,i})/\lambda_{0,i}$ computed relative to the rest wavelengths. For transparency and optional regression testing, the implementation can optionally print diagnostic information, including the half-window used for each line and time step (and, when enabled, fitted parameter values).

Figure~\ref{fig:fig1} illustrates a representative spectrum from the synthetic time series together with the fitted model segments and the recovered line centres. The ground-truth centres for the selected time step are overplotted, enabling direct visual verification of centre recovery across the spectral window. The corresponding per-line residuals at that time step demonstrate sub-picometre accuracy for the adaptive fitting approach under these controlled conditions, with absolute centre residuals of
(1) 0.01\,pm (0.007\,km\,s$^{-1}$),
(2) 0.10\,pm (0.103\,km\,s$^{-1}$),
(3) 0.05\,pm (0.049\,km\,s$^{-1}$),
(4) 0.01\,pm (0.008\,km\,s$^{-1}$),
(5) 0.03\,pm (0.033\,km\,s$^{-1}$),
(6) 0.03\,pm (0.029\,km\,s$^{-1}$),
(7) 0.01\,pm (0.005\,km\,s$^{-1}$),
(8) 0.02\,pm (0.018\,km\,s$^{-1}$),
(9) 0.01\,pm (0.009\,km\,s$^{-1}$),
and (10) 0.02\,pm (0.019\,km\,s$^{-1}$).

In practical use, the most dataset-dependent settings are the expected rest wavelengths and the per-line minimum/maximum half-window bounds, which should reflect the spectral sampling, typical line widths, and the degree of crowding in the observed window. By contrast, the overall logic of the method -- line-guided seeding, close-pair ownership control, bounded fitting, iterative recentering, and morphology-aware handling -- is portable and does not require substantial redesign between datasets. In this sense, \texttt{LineFit} is intended to be tuned mainly at the level of conservative per-line bounds, rather than through highly case-specific heuristics.

\subsection{Reproducibility and code availability}
\label{subsec:code_availability}

The line-fitting implementation (\texttt{LineFit}) is distributed as part of the workflow repository \texttt{WaLSAlib}\footnote{\url{https://GitHub.com/WaLSAteam/WaLSAlib}}, which collects practical routines that connect calibrated data to analysis-ready time series for wave and dynamics studies. All Python scripts and notebooks required to reproduce the synthetic testbed, all figures in this paper, and the method comparisons are provided under \texttt{examples/LineFit\_Frontiers} in the same repository.

\subsection{Other line-centre estimators}
\label{subsec:baselines}

We compare the proposed adaptive fitting approach against four commonly used fast estimators. All baseline methods are applied independently per line, using a per-line half-window in index space: we use the same nominal per-line half-window settings as for \texttt{LineFit} across all baselines for a fair comparison; \texttt{LineFit} may adapt within its prescribed bounds, while the baselines keep the window fixed. For each method, the output is a line-centre estimate $\hat{\lambda}_{i}(t)$ and residuals are evaluated relative to the ground-truth line-centre wavelength $\lambda_i(t)$.

\subsubsection{Parabolic minimum (quadratic interpolation)}
\label{subsec:parabola}

The parabolic-minimum estimator identifies the local intensity minimum within a prescribed search window and fits a quadratic polynomial through the three samples surrounding that minimum. The vertex of the quadratic provides a sub-sample line-centre estimate. This method is computationally inexpensive and often robust for isolated, near-symmetric lines, but it can be biased or destabilised when the profile becomes asymmetric, multi-lobed, or strongly blended.

\subsubsection{Centre-of-gravity (COG) of the absorption depression}
\label{subsec:cog}

The COG estimator is computed from the absorption depression,
\begin{equation}
D(\lambda,t) = 1 - \frac{I(\lambda,t)}{I_{{\rm cont},i}(t)} \ ,
\end{equation}
where $I_{{\rm cont},i}(t)$ is a local continuum estimate for line $i$ at time $t$, obtained robustly from the top fraction of samples within the fitting window (here the top 20\% in each window). The centre is then,
\begin{equation}
\hat{\lambda}_{\mathrm{COG}}(t)=\frac{\sum_\lambda \lambda\,D(\lambda,t)}{\sum_\lambda D(\lambda,t)} \ ,
\end{equation}
where the sums are taken over the wavelength samples within the fitting window, and $D\ge 0$ enforced by clipping. COG is inexpensive and often stable for moderately deformed profiles, but it can be sensitive to blends, window definition (i.e. window ``leakage''), and changes in wing asymmetry because it integrates over the full window.

\subsubsection{Fourier-phase centroid ($k=1$ phase)}
\label{subsec:fftphase}

The Fourier-phase estimator computes a reference-free centroid from the phase of the first Fourier component ($k=1$) of an apodised (Hanning) depression profile within the fitting window. Conceptually related to Fourier-tachometer style Doppler proxies, this estimator can be efficient and less sensitive to pixel-scale noise, but can fail or become biased when the profile contains multiple competing structures within the window (e.g. split cores or strong blends).

\subsubsection{Weighted polynomial core fit (Poly6w)}
\label{subsec:poly6w}

The weighted polynomial method fits a high-order polynomial (here order 6) to a narrow core region around the local minimum, using weights that emphasise samples closest to the core (inverse-distance weighting). The centre is obtained by locating the minimum of the fitted polynomial within the fit interval (e.g. via stationary points of the derivative, with interval-endpoint checks). This can outperform simple parabolic interpolation for smooth symmetric cores, but remains vulnerable to multi-lobed cores and to mis-centring in crowded windows.

\section{Analysis and Results}
\label{sec:results}

We quantify estimator performance in two complementary ways. First, we evaluate instantaneous line-centre errors at a representative time step (Figure~\ref{fig:fig2}), where the truth centres are known exactly. Second, we assess time-series fidelity over the full 50\,min sequence (300 time steps) by comparing recovered LOS-velocity time series against the truth $v_i(t)$ in Figure~\ref{fig:fig3}, then by propagating the resulting time series into standard wave diagnostics (Figure~\ref{fig:fig4}).

\subsection{Single-time-step centre accuracy in the spectral window}
\label{subsec:results_frame}

\begin{figure*}
\centering
\includegraphics[width=\textwidth]{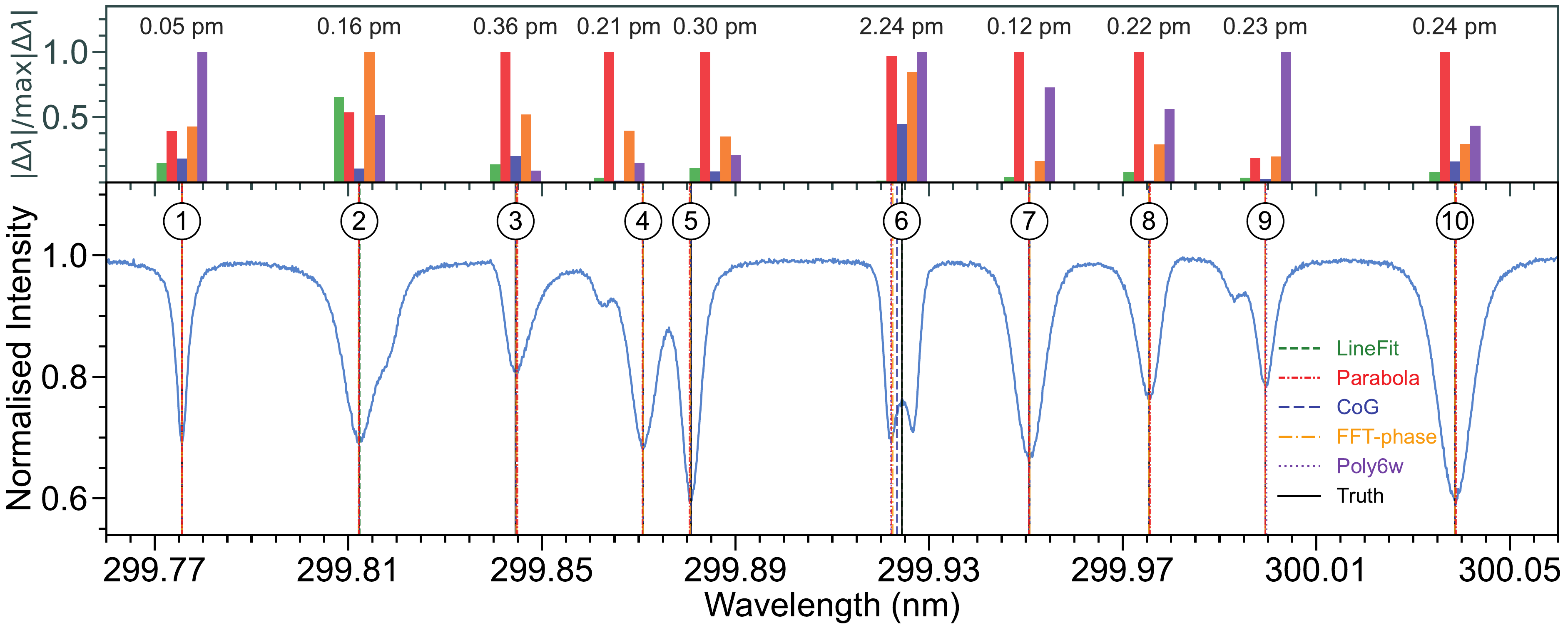}
\caption{Instantaneous centre accuracy comparison for a representative synthetic time step ($t=0$\,s). Lower panel: spectrum with recovered centres from \texttt{LineFit} and four fast baselines overplotted, alongside the truth centres. Upper panel: normalised absolute centre residuals ($|\Delta\lambda_i|/\max|\Delta\lambda_i|$) across methods for each line; text annotations show the corresponding absolute maximum $|\Delta\lambda_i|$ per line (pm).}
\label{fig:fig2}
\end{figure*}

Figure~\ref{fig:fig2} compares five line-centre estimators for one time step (at $t=0$\,s) within a window containing ten lines, including a close blended pair (lines 4--5) and a split-core/emission-reversal candidate (line~6). The lower panel shows the spectrum and the recovered centres overplotted alongside the truth centres, while the upper panel summarises the absolute residuals $|\Delta\lambda_i| = |\hat{\lambda}_i-\lambda_i|$ for each line, normalised by the maximum $|\Delta\lambda_i|$ across methods for that line to highlight relative method performance line-by-line. The label above each line group gives the absolute maximum $|\Delta\lambda_i|$ in picometers across methods for that line.

For this example time step, \texttt{LineFit} achieves sub-picometre centre recovery for all ten lines, with a maximum absolute residual of $0.103$\,pm across the set. The fast baseline estimators show small errors for many isolated lines, but exhibit substantially larger excursions for the split-core line (line~6), reaching picometre-level offsets ($|\Delta\lambda_i|\approx 1.0$--$2.2$\,pm depending on method; $|\Delta\lambda_i|=0.03$\,pm for \texttt{LineFit}). This behaviour is consistent with the known failure modes of minimum- and core-centroid proxies under multi-lobed cores: the inferred ``centre'' can shift towards one lobe or towards a blended feature when the core morphology becomes non-single-valued within the window. By contrast, in \texttt{LineFit} the morphology-aware handling of the split-core case is intended precisely to avoid such lobe-hopping, so that the recovered centre remains tied to the overall core structure rather than to one of the instantaneous lobes.

\subsection{Time-series fidelity: velocity residuals over 50\,min}
\label{subsec:results_timeseries}

Instantaneous accuracy does not guarantee time-series stability: rare centre jumps can inject broadband power and bias wave diagnostics. We therefore compute, for each line and each method, the LOS velocity time series following,
\begin{equation}
\hat{v}_i(t) = c\,\frac{\hat{\lambda}_i(t)-\lambda_{0,i}}{\lambda_{0,i}} \ ,
\end{equation}
and evaluate the residuals against the truth $v_i(t)$ over all 300 time steps. As a summary metric we use the per-line Root Mean Square Error (RMSE),
\begin{equation}
\mathrm{RMSE}_i = \sqrt{\langle\left(\hat{v}_i(t)-v_i(t)\right)^2\rangle_t} \ ,
\end{equation}
where $\langle\cdot\rangle_t$ denotes averaging over time samples; in practice, the RMSE is computed over time samples where both truth and estimate are finite.

\begin{figure*}
\centering
\includegraphics[width=\textwidth]{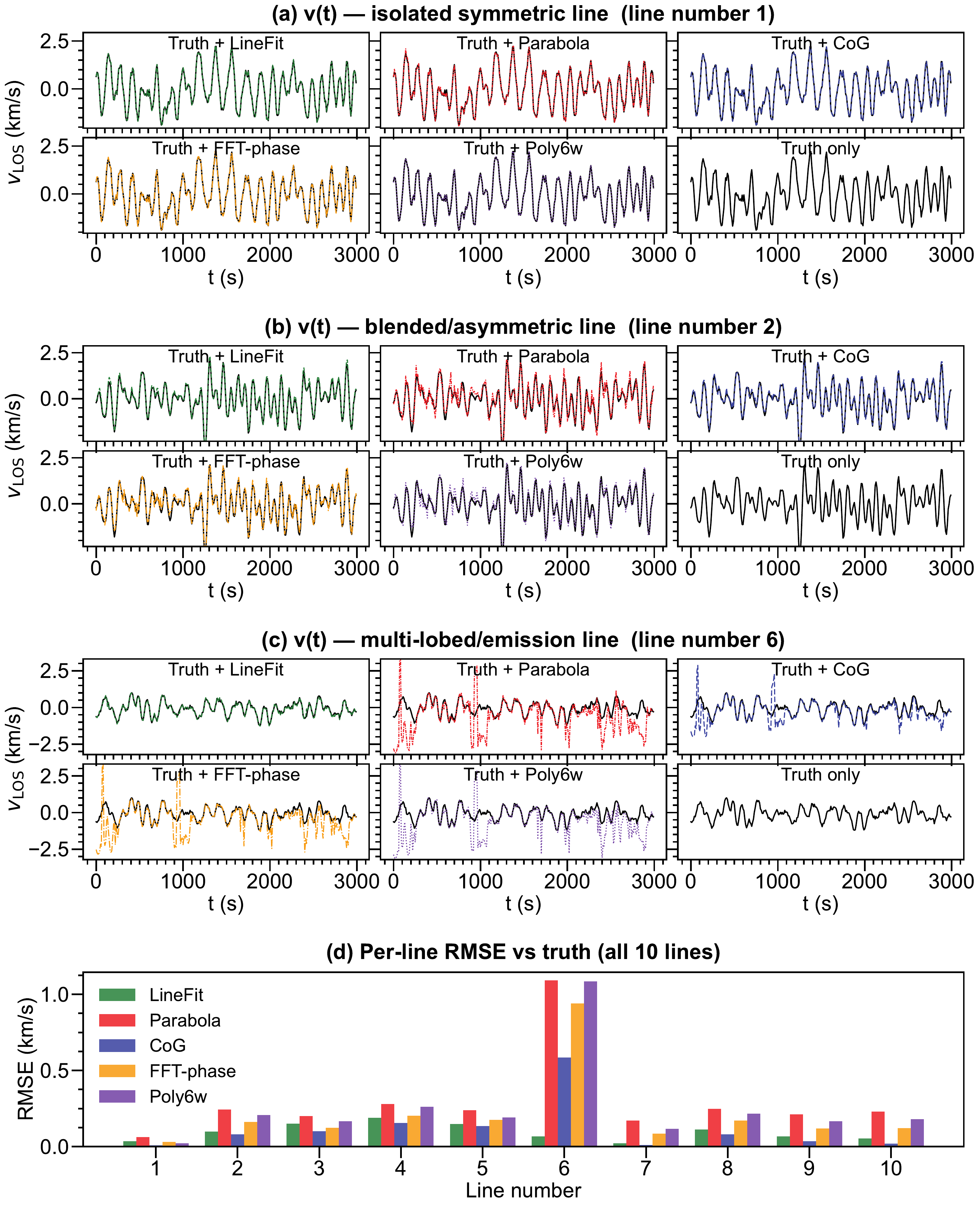}
\caption{Velocity time-series fidelity over the full synthetic sequence. Rows (a--c): examples for an isolated symmetric line (line~1), an asymmetric/blended line (line~2), and a split-core/emission-reversal line (line~6). Each row shows truth and one recovered method per sub-panel to aid visual comparison. Residual-versus-truth time-series plots for the same representative lines are provided in Supplementary Figure~S1.} Panel (d): per-line RMSE of $\hat{v}(t)$ against the truth for all ten lines and all methods.
\label{fig:fig3}
\end{figure*}

Figure~\ref{fig:fig3}(a--c) shows representative velocity time-series comparisons for three lines spanning increasing profile complexity: an isolated symmetric line (line~1), an asymmetric/blended line (line~2), and the split-core/emission-reversal line (line~6). For the isolated case, all methods track the truth reasonably well, typically showing the closest agreement and the smallest residual structure. For the asymmetric/blended case, the dispersion between methods increases, consistent with sensitivity to evolving asymmetry and to blends or window leakage (when a fixed window includes neighbouring structure); \texttt{LineFit} and COG broadly capture the dominant variation, whereas FFT-phase and, more prominently, the parabolic minimum and Poly6w exhibit larger deviations at several intervals.

The most pronounced differences occur for the split-core/emission-reversal line (line~6). Here \texttt{LineFit} remains stable and closely follows the truth throughout the sequence, while all four fast estimators exhibit substantially larger mismatches, including episodes of divergence associated with time steps where the core morphology becomes strongly multi-lobed (see the spectrum evolution in the online movie\footnote{\url{https://WaLSA.team/LineFit/synthetic_spectrum_timeseries}}, which can be regenerated from the public repository). This behaviour is reflected directly in the per-line RMSE summary (Figure~\ref{fig:fig3}d): for line~6, \texttt{LineFit} yields $\mathrm{RMSE}=0.067$\,km\,s$^{-1}$, whereas the fast methods range from $\mathrm{RMSE}\approx 0.58$\,km\,s$^{-1}$ (COG) to $\approx 1.09$\,km\,s$^{-1}$ (parabolic minimum), i.e. nearly an order-of-magnitude degradation in time-series fidelity for this complex-core case.

Across all ten lines, \texttt{LineFit} provides the best overall performance in this testbed (mean RMSE $0.094$\,km\,s$^{-1}$; median $0.083$\,km\,s$^{-1}$), with COG second-best (mean RMSE $0.121$\,km\,s$^{-1}$).
The parabolic minimum, FFT-phase, and Poly6w methods show larger mean RMSE values (0.30, 0.21, and 0.26\,km\,s$^{-1}$, respectively), driven primarily by their failures on the split-core line and, to a lesser extent, by sensitivity to evolving asymmetry and blends. We emphasise that these baseline implementations are intentionally kept simple and uniform across lines to provide a controlled comparison; more aggressive method-specific tuning and case-dependent heuristics may improve performance on some profiles, at the expense of additional parameter complexity and reduced portability.

Whether a given centre error is scientifically acceptable depends primarily on how it propagates into the derived LOS-velocity time series and wave diagnostics. In this benchmark, the simpler estimators are often adequate for well-behaved lines, but become problematic in the morphologically complex stress cases that motivate \texttt{LineFit}.

\subsection{Impact on wave diagnostics: refined global wavelet spectra}
\label{subsec:results_wavelet}

To quantify how centre-extraction differences propagate into wave inference, we compute Refined Global Wavelet Spectra (RGWS) from each recovered velocity time series using the WaLSAtools\footnote{\url{https://GitHub.com/WaLSAteam/WaLSAtools}} \citep{2025NRvMP...5...21J, walsatools..2025...17569951} implementation. Briefly, RGWS is the time-integrated Morlet wavelet power refined by retaining only statistically significant power (95\% confidence) and excluding regions within the cone of influence (subject to edge effects).
All velocity signals are linearly detrended and apodised using a Tukey window ($\alpha=0.1$) prior to the wavelet analysis.

\begin{figure*}
\centering
\includegraphics[width=\textwidth]{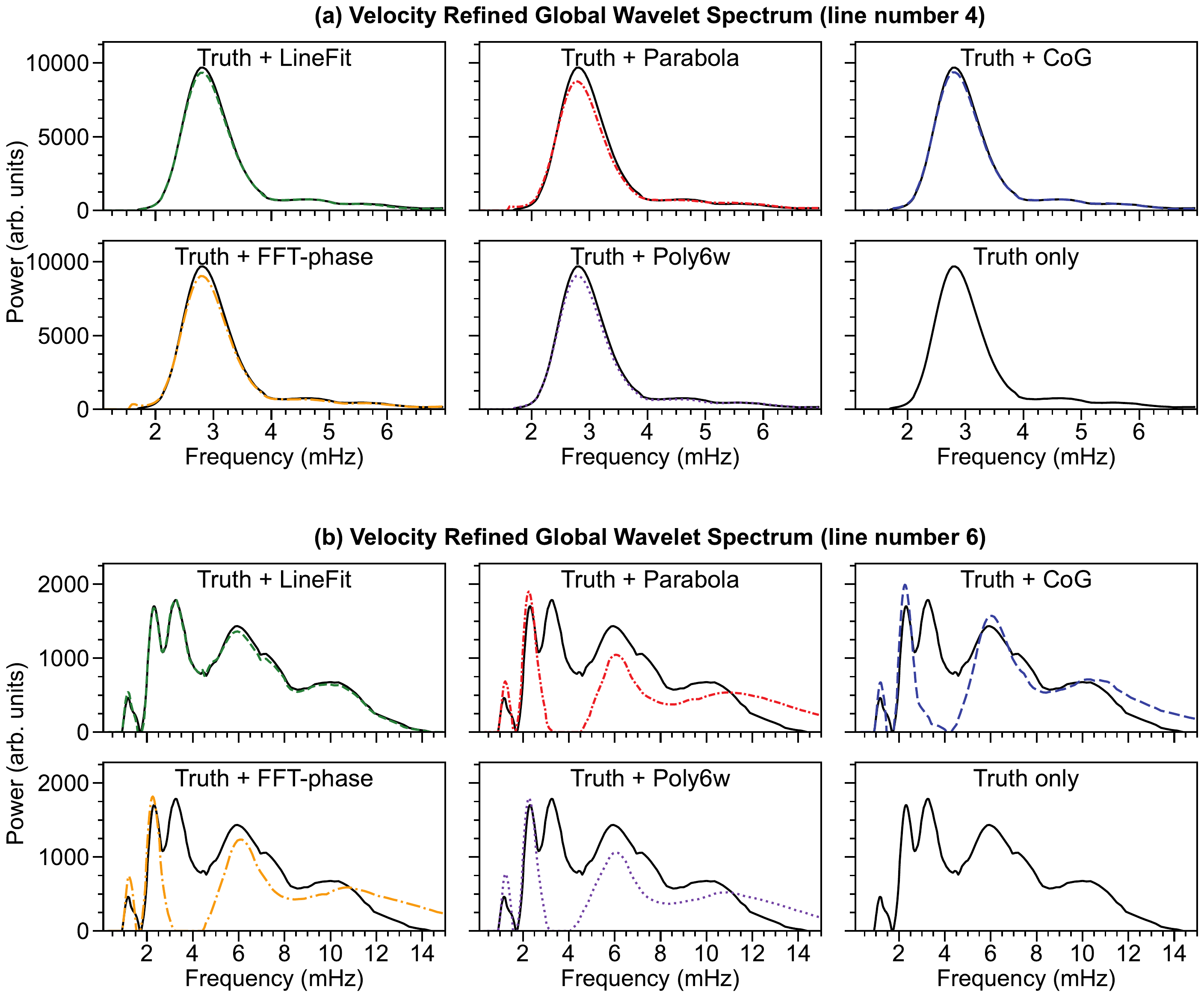}
\caption{Propagation of centre-extraction errors into wave diagnostics. Refined Global Wavelet Spectra (RGWS) computed from velocity time series recovered by each method, compared against the truth RGWS. (a) Example blended line (line~4): all methods recover the main peak structure, with \texttt{LineFit} and COG closest to the truth in peak power. (b) Split-core/emission-reversal line (line~6): only \texttt{LineFit} reproduces the full truth peak structure; fast baselines miss or shift spectral components (notably near $\sim$3.2\,mHz) and show power bias at higher frequencies.}
\label{fig:fig4}
\end{figure*}

Figure~\ref{fig:fig4} compares RGWS for two representative lines: a blended line that is one component of the close pair and exhibits relatively strong depth variability (line~4) and the split-core/emission-reversal line (line~6). For line~4 (which has the second-largest RMSE across the five methods, though still modest), all methods recover consistent peak locations and broadly similar spectral structure, indicating that for moderately complex but single-core profiles the baseline estimators can preserve the dominant oscillation content. However, even in this regime \texttt{LineFit} and COG provide the closest match to the truth spectrum in both peak power and overall spectral shape, while FFT-phase and Poly6w show slightly larger deviations and the parabolic method performs worst in this example.

For line~6, the differences become qualitatively important. \texttt{LineFit} reproduces both the locations and amplitudes of the truth RGWS peaks, with only minor residual mismatch at the highest-frequency components. In contrast, the four fast estimators miss or distort key spectral content: while some peak locations are recovered, their associated power is often biased, and the power enhancement near $\sim$3.2\,mHz is not reproduced. At higher frequencies (e.g. near $\sim$10\,mHz) the spectra show shifted peaks and additional power bias. This outcome is consistent with the time-domain behaviour in Figure\ref{fig:fig3} and demonstrates that centre instabilities in complex-core time steps can translate directly into missing or mislocated oscillation power, thereby biasing subsequent wave analyses, including phase/coherence and mode-identification diagnostics.

\section{Discussion and Conclusions}
\label{sec:conclusions}

We have presented an adaptive multi-line fitting methodology (\texttt{LineFit}) aimed at producing stable line-core intensity and LOS-velocity time series from dense spectral windows in which line profiles evolve rapidly and nonlinearly. The motivation is practical but scientifically consequential: discontinuities and occasional misidentification of the line core can inject artefacts into time-series diagnostics, bias dominant-frequency estimates, and possibly degrade phase/coherence measurements that underpin MHD-wave inferences. This challenge is becoming increasingly prominent as modern solar spectrographs and IFUs (e.g. DKIST/DL-NIRSP and ViSP, {\sc Sunrise iii}/SUSI and SCIP, and FRANCIS) routinely deliver crowded windows containing many lines per spatial sample and per time step.

To isolate performance in a controlled and fully reproducible way, we benchmarked \texttt{LineFit} using a realistic synthetic near-UV time series with unambiguous ground truth for the time-dependent line centres. The synthetic testbed was constructed to explicitly stress the main failure modes encountered in practice: evolving asymmetries, close blends, and intermittent split-core (reversal) morphologies in which a core brightening appears relative to the surrounding absorption depression. Against four widely used fast estimators (parabolic minimum, COG of absorption depression, Fourier-phase centroid, and a weighted polynomial core fit), \texttt{LineFit} consistently recovered accurate centres in representative time steps (Figure~\ref{fig:fig2}), while also producing the most robust LOS-velocity time series in the presence of multi-lobed cores (Figure~\ref{fig:fig3}). Across the full time series, several of the fast methods performed competitively for many isolated or moderately deformed lines, and in our testbed the COG estimator in particular was often among the best-performing baselines for simple absorption profiles. However, in the key stress case (an intermittently split-core line with strong core-structure variability) \texttt{LineFit} was the only method that consistently maintained agreement with the truth time series and avoided large excursions in the recovered velocity.

A central point of this work is that centre-tracking stability should be assessed not only by per-time-step residuals but also by the downstream wave products used in MHD studies. When the derived velocity contains sporadic mis-tracking events, the resulting time series can acquire artificial power and distort spectral diagnostics. This is clearly demonstrated by the RGWS: for a representative blended but single-core line all methods broadly reproduce the main power enhancements, whereas for the split-core case the fast estimators systematically miss or misplace key spectral peaks and misestimate power compared with the truth, while \texttt{LineFit} preserves both the peak structure and the power distribution, with only minor differences at a few frequencies (Figure~\ref{fig:fig4}). These differences are precisely the kinds of discrepancies that matter for multi-height wave and dynamics studies across the solar atmosphere.

The synthetic benchmark provides the key advantage that the true line centres and velocities are known exactly, allowing direct measurement of centre residuals, velocity residuals, and propagated wave-diagnostic errors. In real observations, such absolute residuals are generally not accessible, and performance must instead be assessed through internal consistency, temporal stability, resistance to centre-hopping, and the plausibility and continuity of downstream products such as velocity spectra and wavelet diagnostics. We therefore view the synthetic tests as a controlled validation benchmark, while the practical value of \texttt{LineFit} in real data is demonstrated primarily by its stable behaviour in dense-window applications where complex and evolving profiles are present. Additional real-data effects such as scattered light, residual flat-field structure, correlated noise, and wavelength-calibration uncertainties may modify the absolute error scale, but they do not change the core motivation of the method, namely to reduce unstable centre tracking in precisely those cases where such instabilities would most strongly bias time-series diagnostics.

While the paper focuses on a synthetic benchmark, \texttt{LineFit} has already been applied to dense-spectral-window observations in real-data studies and contexts: it has been used in the analysis of {\sc Sunrise iii}/SUSI near-UV spectra containing more than 100 spectral lines for active-region wave diagnostics \citep{Jafarzadeh2026_paper1, Jafarzadeh2026_paper2}, and it has been used to extract line-core parameters in a FRANCIS integral-field analysis \citep{2026FrASS...Chambers_inprep}. In such real-data applications, the most relevant indicators are not absolute residuals to a known truth, but whether the extracted line-core products remain temporally stable, physically plausible, and sufficiently robust for downstream wave and dynamics analysis. These applications demonstrate that the method is practical for novel, information-rich datasets and that it integrates naturally into wave-analysis workflows. We note, however, that the occurrence frequency of multi-lobed or reversal-like profiles in real data is strongly dataset-, line-, and target-dependent; such morphologies occur often enough in dynamic solar spectroscopy to warrant explicit handling when stable time-series extraction is required. At the same time, we emphasise that fast estimators remain valuable: for large-volume datasets and for lines that remain single-lobed and isolated, methods such as COG or Fourier-phase proxies can be attractive due to their computational efficiency, and they are used in contemporary analyses where their assumptions are well matched to the data (e.g. Poly6w in \citealt{Lagg+2026}; FFT-phase approaches in \citealt{2026FrASS...Grant_inprep}). Our results therefore support a pragmatic view: robust adaptive fitting is most critical for the subset of lines, locations, and times where the profiles become complex -- often the same regimes that carry the strongest dynamical signatures -- while fast estimators can provide efficient performance elsewhere, provided their limitations are recognised and quality control is applied.

All code required to reproduce the synthetic testbed and figures (including the \texttt{LineFit} implementation) is available via the \texttt{WaLSAlib} repository (see Section~\ref{subsec:code_availability}). The current implementation is intentionally modular and can be extended to include additional profile families, morphology-specific quality flags, and instrument- or line-dependent constraints as required by particular datasets. A natural next step is to export additional shape diagnostics (e.g. proxy widths, asymmetry measures, and core-structure metrics); although width estimation is non-trivial for deep lines whose wings do not return to the local continuum, it can be handled using robust, locally normalised definitions and reported with quality flags.

Looking forward, the increasing scale of next-generation, high-cadence slit-based spectroscopic and IFU datasets makes acceleration strategies increasingly important. As a proof-of-principle, we trained a lightweight supervised convolutional neural network (CNN; \citealt{1998IEEEP..86.2278L, Goodfellow2016}) emulator on the synthetic testbed, using \texttt{LineFit} outputs as training labels (ground truth was withheld from training and used only for verification). The emulator predicts the residual line-centre shift relative to a robust coarse centre computed from each spectrum, reducing the dynamic range the network must learn and decoupling coarse localisation from fine-structure refinement. A hybrid inference mode routes only morphologically complex profiles -- identified by a split-core/multi-peak morphology flag and a Monte Carlo dropout uncertainty estimate -- to \texttt{LineFit} as a fallback, while the emulator handles the remainder. On the reserved validation set (i.e. the final 20\% of the time series, strictly withheld from training), the emulator achieves a median absolute centre error of 0.049\,pm and a 95th-percentile error of 0.245\,pm relative to the ground truth, with 87\% of validation samples agreeing with \texttt{LineFit} to within 0.1\,pm. The hybrid fallback rate is 2.1\% of all samples (applied only to the stress-case line), and wave diagnostics computed from the emulator and hybrid outputs are well preserved relative to the \texttt{LineFit} reference. The hybrid speed-up relative to running \texttt{LineFit} across the full sequence is approximately three orders of magnitude on the hardware used here, with the emulator component alone approaching five orders of magnitude; the operationally relevant hybrid figure accounts for the \texttt{LineFit} fallback cost on the small flagged fraction (specific numbers and timing methodology are given in Supplementary Section~\ref{supp:emulator}). Such a hybrid strategy can preserve the physical transparency and traceability of parametric fitting while achieving the throughput required for very large field-of-view and high-cadence time-series datasets. This proof-of-principle is intentionally limited in scope, and full development -- including larger training sets, cross-validation, and instrument-specific tuning -- is left for future work. More broadly, we expect that robust, adaptive extraction of line-core products will remain a critical enabling step for wave and dynamics studies using dense-window spectroscopy, particularly when multi-line diagnostics are used to infer energy transport and mode coupling across atmospheric layers.

\section*{Conflict of Interest Statement}
The authors declare that the research was conducted in the absence of any commercial or financial relationships that could be construed as a potential conflict of interest.

\section*{Author Contributions}
SJ, DBJ, MS, PHK, and GC contributed to the conceptualisation of the study and the design of the validation strategy. SJ led the software development, implemented the core \texttt{LineFit} methodology and synthetic testbed, performed the formal analysis, and produced the figures and benchmark products. DBJ, MS, PHK, GC, SDTG, MB, and TJD contributed to software development, validation, and testing, with DBJ, PHK, GC, and SDTG also supporting the visualisation and interpretation of results. All authors contributed to manuscript writing and reviewed the final text.

\section*{Funding}
UK Science and Technology Facilities Council (STFC) consolidated grants ST/T00021X/1 and ST/X000923/1.
UK STFC PATT Travel Grant UKRI372.
Leverhulme Trust Research Project Grant RPG-2019-371. 
UK Space Agency National Space Technology Programme grant SSc-009.
Research Council of Norway project no. 262622.
The Royal Society award no. Hooke18b/SCTM.
NASA grants 19-HSODS-004 and 21-SMDSS21-0047.
International Space Science Institute (ISSI) Team 502.

\section*{Acknowledgments}
SJ, DBJ, and GC wish to thank the UK Science and Technology Facilities Council (STFC) for the consolidated grants ST/T00021X/1 and ST/X000923/1, alongside the PATT Travel Grant UKRI372.
DBJ acknowledges support from the Leverhulme Trust via the Research Project Grant RPG-2019-371. 
DBJ and SDTG also acknowledge funding from the UK Space Agency via the National Space Technology Programme (grant SSc-009).
SJ acknowledges support from the Rosseland Centre for Solar Physics, University of Oslo, Norway and the Max Planck Institute for Solar System Research, Germany.
MB acknowledges that this publication was produced while attending the PhD programme in PhD in Space Science and Technology at the University of Trento, Cycle XXXIX, with the support of a scholarship financed by the Ministerial Decree no. 118 of 2nd March 2023, based on the NRRP - funded by the European Union - NextGenerationEU - Mission 4 ``Education and Research'', Component 1 ``Enhancement of the offer of educational services: from nurseries to universities'' - Investment 4.1 ``Extension of the number of research doctorates and innovative doctorates for public administration and cultural heritage'' - CUP E66E23000110001.
We wish to acknowledge scientific discussions with the Waves in the Lower Solar Atmosphere (WaLSA; \href{https://WaLSA.team}{www.WaLSA.team}) team, which has been supported by the Research Council of Norway (project no. 262622), The Royal Society \citep[award no. Hooke18b/SCTM;][]{2021RSPTA.37900169J}, and the International Space Science Institute (ISSI) in Bern through ISSI International Team project 502 ``WaLSA: Waves in the Lower Solar Atmosphere at High Resolution''.

\section*{Data Availability Statement}
All datasets analysed in this study are synthetic and were generated using a publicly available code. The synthetic testbed generator, the \texttt{LineFit} implementation, and all scripts/notebooks required to reproduce the figures and results are available in the \texttt{WaLSAlib} repository (\url{https://GitHub.com/WaLSAteam/WaLSAlib}), under \texttt{examples/LineFit\_Frontiers}, where the synthetic data cube is also provided as FITS format.

\bibliographystyle{Frontiers-Harvard}
\bibliography{article, bass}

\clearpage
\markboth{}{}
\setcounter{page}{1}
\vspace*{-2.5cm}
\def\firstAuthorLast{}

\vspace*{0.5cm}

{\noindent\huge\bfseries Supplementary Material}

\vspace*{0.5cm}

\setcounter{section}{0}
\setcounter{subsection}{0}
\setcounter{table}{0}
\setcounter{figure}{0}
\setcounter{equation}{0}
\renewcommand{\thesection}{S\arabic{section}}
\renewcommand{\thesubsection}{S\arabic{section}.\arabic{subsection}}
\renewcommand{\thetable}{S\arabic{table}}
\renewcommand{\thefigure}{S\arabic{figure}}
\renewcommand{\theequation}{S\arabic{equation}}

\section{Residual time-series comparison}
\label{supp:figS1_residuals}

To complement Figure~\ref{fig:fig3} of the main text, we show in Supplementary Figure~\ref{fig:residuals} the residual LOS-velocity time series, $\hat{v}(t)-v_{\rm truth}(t)$, for the same three representative lines: an isolated symmetric line, a blended/asymmetric line, and the split-core/emission-reversal stress-case line. Presenting the results as residuals relative to the truth makes the time-dependent deviations of each method easier to compare directly than in the overplotted velocity time series of the main text. This residual representation therefore illustrates directly why the error metrics and downstream wave diagnostics favour \texttt{LineFit} in the challenging cases that motivate the method.

\begin{figure*}[h]
\centering
\includegraphics[width=1.0\textwidth]{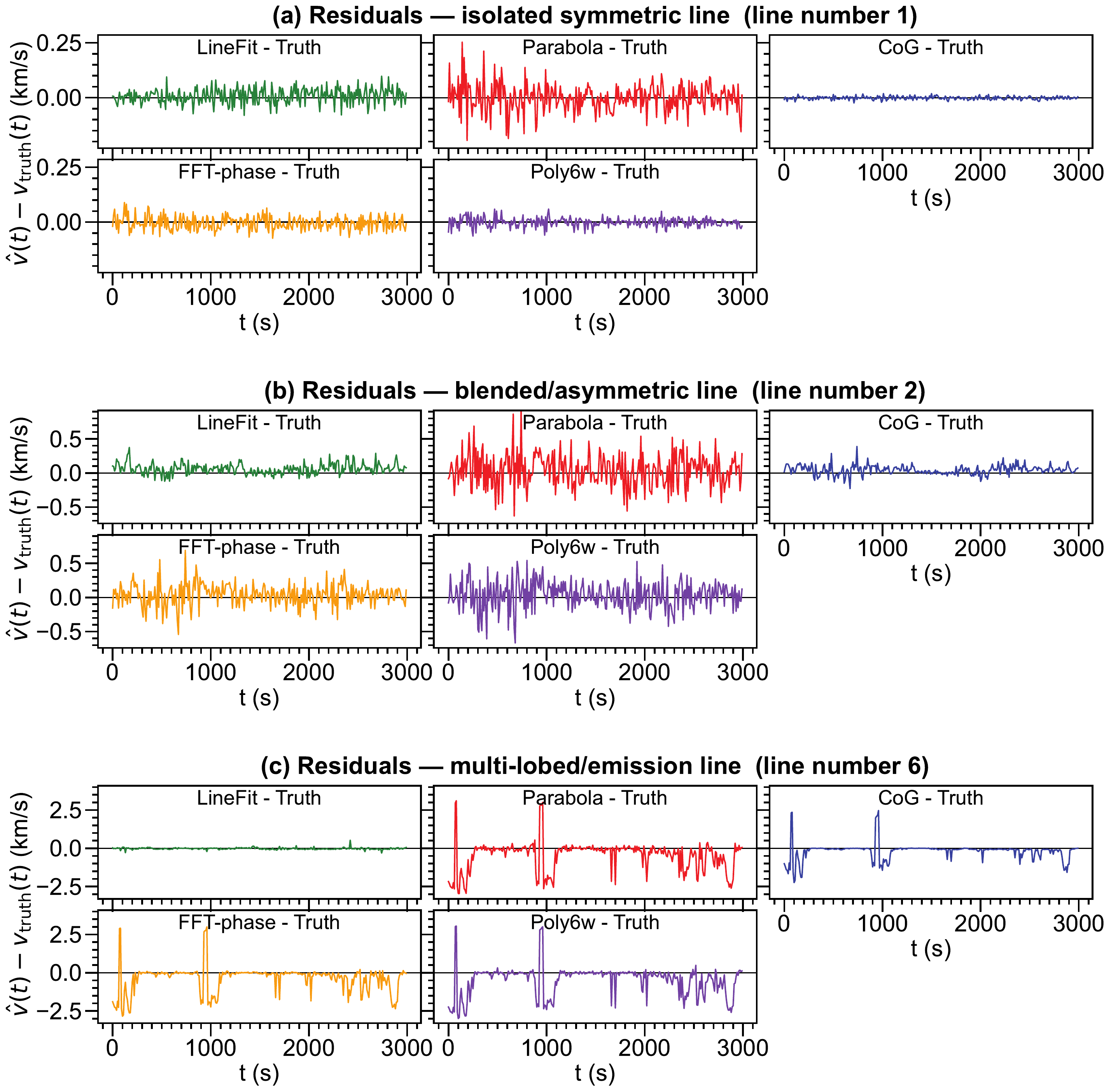}
\caption{Residual LOS-velocity time series, $\hat{v}(t)-v_{\rm truth}(t)$, for the same three representative lines shown in Figure~\ref{fig:fig3} of the main text: (a) an isolated symmetric line, (b) a blended/asymmetric line, and (c) a multi-lobed/emission-reversal line. Each mini-panel shows one method relative to the truth with the same y-scale to allow direct visual comparison of the time-dependent deviations.}
\label{fig:residuals}
\end{figure*}

\section{Proof-of-principle hybrid acceleration via supervised emulation}
\label{supp:emulator}

To illustrate a practical route towards scalable processing while retaining the robustness of parametric fitting, we implemented a proof-of-principle (PoP) hybrid acceleration scheme using a small supervised convolutional neural network (CNN) emulator. The intent is to demonstrate feasibility on the same synthetic testbed used throughout the paper; full development is deferred to future work.

\subsection{Setup and training}

The emulator takes, for each (time step, line) pair, a normalised intensity-inverted local spectral window centred on a robust coarse centre -- plus a binary validity mask distinguishing true spectral samples from edge-replicated padding used near the spectral-window boundaries -- and predicts the residual line-centre shift relative to that coarse centre. \texttt{LineFit} outputs serve as training labels; ground truth from the synthetic generator is used only for evaluation (not for training, calibration, or threshold selection). A per-line embedding allows a single network to serve all ten lines. Training uses a strictly time-ordered split to prevent temporal leakage: the first 80\% of the time series (240/300 steps) is used for training and calibration, with the last 10\% of that training window reserved for calibration (24 steps) and the remaining 216 steps used for gradient updates; the final 20\% (60 steps) is reserved for validation. Training stopped automatically after 42 epochs using early stopping: the validation loss did not improve for 16 consecutive epochs (the patience setting), so optimisation was halted to avoid overfitting. This behaviour is expected and indicates that additional epochs would not have improved generalisation for this configuration. All fixed hyperparameters are available in the public repository (under \texttt{examples/LineFit\_Frontiers}) alongside the emulator script.

Hybrid inference is applied only to the stress-case line (line~6). A morphology flag is raised on split-core-like or multi-peak profiles; for flagged samples, 30 Monte Carlo dropout passes provide an uncertainty estimate. A threshold calibrated on morphology-flagged calibration samples (the 95th percentile of the dropout standard deviation, 0.143\,pm) determines which flagged samples are routed to \texttt{LineFit}. In this PoP the fallback uses pre-computed \texttt{LineFit} outputs; in production, only the flagged time steps would invoke \texttt{LineFit} live.

The morphology criteria are intentionally conservative. In borderline cases, a false-positive classification mainly increases the fallback rate and therefore reduces the speed-up, but does not compromise the fitted result because the sample is simply routed to \texttt{LineFit}. By contrast, a false negative would leave a difficult profile to the emulator; in the present hybrid scheme this risk is reduced by the additional Monte Carlo dropout uncertainty gate, so that morphology and uncertainty act as complementary safeguards rather than as a single hard classifier.

\subsection{Accuracy}

\begin{figure*}[!h]
\centering
\includegraphics[width=1.0\textwidth]{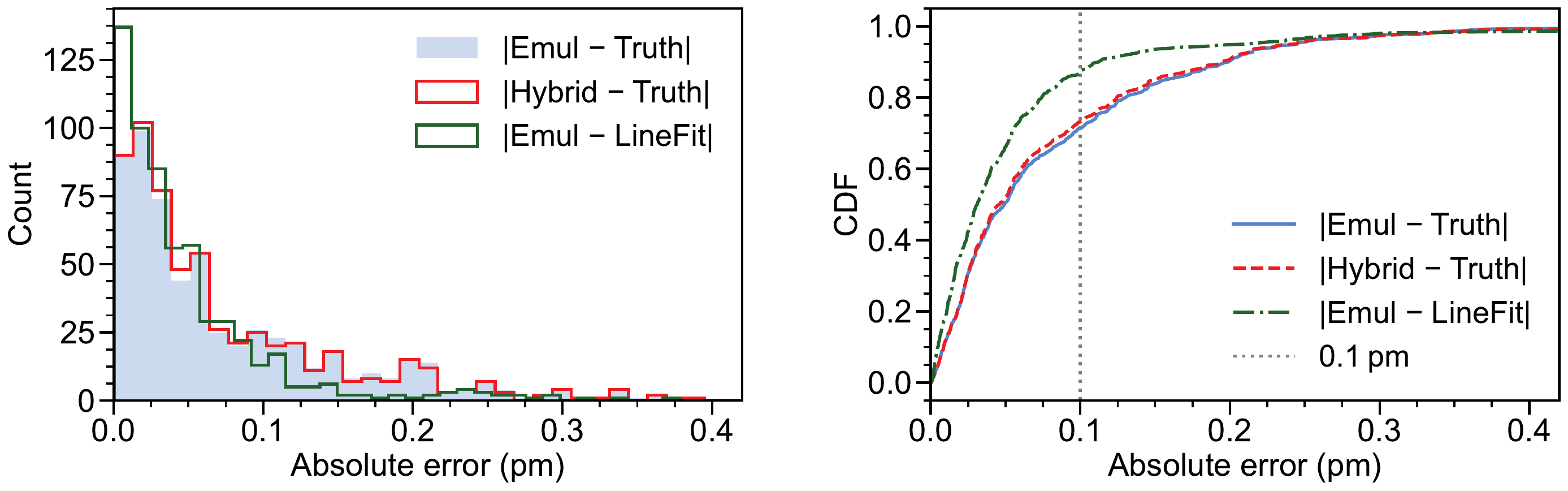}
\caption{Validation-set absolute centre error distributions for the CNN emulator and hybrid scheme (both vs ground truth) and for the emulator vs \texttt{LineFit}. Left: histograms; right: CDFs. The CDFs directly show that 87\% of validation samples agree with \texttt{LineFit} to within 0.1\,pm.}
\label{fig:supp_val_errors}
\end{figure*}

On the validation set the emulator achieves a mean absolute error (MAE) of 0.080\,pm, RMSE\,=\,0.118\,pm, and a 95th-percentile absolute centre error of 0.245\,pm relative to the ground truth (MAE\,=\,0.055\,pm and P95\,=\,0.211\,pm relative to \texttt{LineFit}), with 87\% of samples agreeing with \texttt{LineFit} to within 0.1\,pm. Calibration-set accuracy is nearly identical (P95\,=\,0.228\,pm), suggesting no strong overfitting to the training window. The hybrid fallback fraction is 2.1\%. Because fallback is applied only to the stress-case line, a 2.1\% fallback rate corresponds to 62 of 3000 (time step, line) samples. Figure~\ref{fig:supp_val_errors} shows the validation error distributions as histograms and cumulative distributions (CDFs); the emulator and hybrid curves are nearly coincident at this fallback rate, and the emulator-vs-\texttt{LineFit} distribution is slightly narrower because it captures only the label-to-emulator discrepancy without the residual \texttt{LineFit}-to-truth mismatch. 

\subsection{Wave-diagnostic preservation}

\begin{figure*}[!h]
\centering
\includegraphics[width=0.8\textwidth]{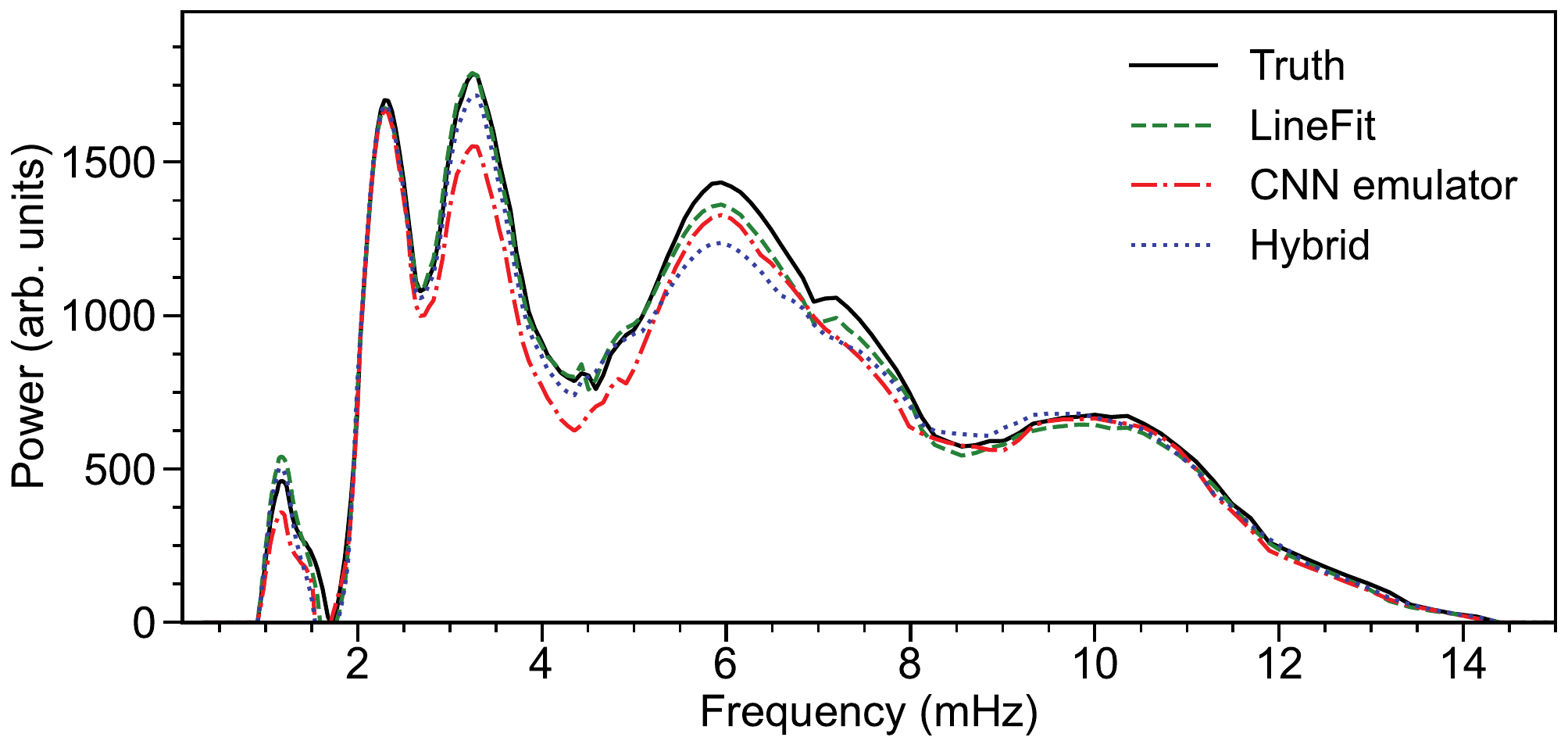}
\caption{RGWS of the LOS velocity time series for the stress-case line (line~6) from \texttt{LineFit}, the CNN emulator, and hybrid approach, compared with the truth.}
\label{fig:supp_rgws}
\end{figure*}

Figure~\ref{fig:supp_rgws} shows the RGWS of the velocity time series for the stress-case line from each method, compared with the truth. \texttt{LineFit} and the hybrid follow the truth most closely across the full frequency range, with some amplitude deviations in the higher frequency regime ($\gtrsim$6\,mHz). The CNN emulator agrees well at frequency peaks but with mild deviations at various locations, consistent with its larger per-step errors at the most extreme morphological episodes. The hybrid, which substitutes \texttt{LineFit} values for 2.1\% of stress-line samples, matches the truth spectral structure more closely than the emulator alone, confirming that the gating logic is effectively targeting the time steps where the emulator diverges most.

\subsection{Computational cost}

Both timing estimates follow the same approach: a representative subset of time steps is timed and the result scaled to the full sequence. The \texttt{LineFit} reference runtime ($\approx$53000\,s, or approximately $\sim$15~hours, for 300 time steps and 10 lines) was estimated by timing 10 uniformly sampled steps. The hybrid fallback cost was estimated by timing 10 uniformly sampled fallback steps for the stress-case line only, yielding a per-step cost of $\approx$0.68\,s per fallback time step (stress line only) and an extrapolated fallback total of $\approx$42\,s for the 62 flagged steps. The emulator inference over all 3000 samples takes $\approx$0.55\,s, giving a hybrid total of $\approx$43\,s and a speed-up of approximately three orders of magnitude relative to the full \texttt{LineFit} run. The emulator-only speed-up approaches five orders of magnitude. All runtime estimates are hardware-dependent (measured on a MacBook Pro with Apple M1 Pro, 8-core CPU, and 16 GB RAM; macOS 26.4) and should be taken as indicative of the scale of the potential acceleration rather than as absolute benchmarks. Nevertheless, with the MAE of 0.080\,pm corresponding to a mean absolute Doppler velocity uncertainty of $\approx$80\,m{\,}s$^{-1}$, combined with a performance improvement of at least three orders of magnitude (i.e. on the order of 15~hours $\rightarrow$ 43~seconds), the hybrid emulation of the \texttt{LineFit} software offers a strong platform for future development and wave studies, especially considering the RGWS spectrum accurately reflects the ground truth wave amplitudes within the commonly observed frequency range of $\sim 3 - 5$\,mHz. 

We did not carry out a systematic CPU benchmark of the four fast baseline estimators in the present study, but these simple window-based methods are expected to remain substantially cheaper than \texttt{LineFit}. The practical point of the PoP emulator/hybrid experiment is therefore not to compete with the absolute speed of the simplest baselines, but to show that a workflow anchored to \texttt{LineFit}-quality labels can move much closer to fast-estimator throughput while retaining a robust fallback for difficult profiles.

\end{document}